\documentclass{svmult}

\usepackage{url}             
\usepackage{times}           
\usepackage{makeidx}         
\usepackage{graphicx}        
\usepackage{multicol}        
\usepackage[bottom]{footmisc}

\makeindex             

\begin{document}

\title*{Traders imprint themselves by adaptively updating their own avatar}
\author{Gilles Daniel\inst{1}\and
Lev Muchnik\inst{2}\and
Sorin Solomon\inst{3}
}
\institute{School of Computer Science, University of Manchester, UK \\
\texttt{gilles@cs.man.ac.uk}
\and Department of Physics, Bar Ilan University, Ramat Gan, Israel \\
\texttt{LevMuchnik@gmail.com}
\and Racah Institute of Physics, Hebrew University of Jerusalem and \\
Lagrange Laboratory for Excellence in Complexity, ISI Foundation, Torino
\texttt{sorin@cc.huji.ac.il}
}
%
%
\maketitle

Simulations of artificial stock markets were considered as early as 1964 \cite{stigler} and multi-agent ones were introduced as early as 1989 \cite{markowitz}. Starting the early 90's \cite{asm,levy,terna1}, collaborations of economists and physicists produced increasingly realistic simulation platforms. Currently, the market stylized facts are easily reproduced and one has now to address the realistic details of the Market Microstructure and of the Traders Behaviour. This calls for new methods and tools capable of bridging smoothly between simulations and experiments in economics. 

We propose here the following \emph{Avatar-Based Method} (ABM). The subjects implement and maintain  their Avatars (programs encoding their personal decision making procedures) on NatLab, a market simulation platform. Once these procedures are fed in a computer edible format, they can be operationally used as such without the need for belabouring, interpreting or conceptualising them. Thus ABM short-circuits the usual behavioural economics experiments that search for the psychological mechanisms underlying the subjects’ behaviour. Finally, ABM maintains a level of objectivity close to the classical behaviourism while extending its scope to subjects' decision making mechanisms.

We report on experiments where Avatars designed and maintained by humans from different backgrounds (including real traders) compete in a continuous double-auction market. Instead of viewing this as a collectively authored computer simulation, we consider it rather as a new type of computer aided experiment. Indeed we consider the Avatars as a medium on which the subjects can imprint and refine interactively representations of their internal decision making processes. Avatars  can be objectively validated (as carriers of a faithful replica of the subject decision making process) by comparing their actions with the ones that the subjects would take in similar situations. We hope this unbiased way of capturing the adaptive evolution of real subjects behaviour may lead to a new kind of behavioural economics experiments with a high degree of reliability, analysability and reproducibility.

\section{Introduction}

In the last decade, generic stylized facts were reproduced with very simple agents by a wide range of models \cite{bak1,stanley1,lux,cont2,slanina1,bouchaud}. By the very nature of their generic properties, those models teach us little on real particular effects taking place as result of real particular conditions within the market. In order to understand such specific market phenomena, one may need to go beyond "simple-stupid" traders behaviour \cite{axelrod2}. Thus the task of the present generation of models is to describe and explain the observed collective market phenomena in terms of the actual behaviour of the individuals.

For a long while, classical economics assumed individuals were homogeneous and behaved rationally. Thus it was not necessary to study real people behaviour since (presumably) there is only one way to be rational. Even after the conditions of rationality and homogeneity were relaxed, many models did it by postulating arbitrary departures not necessarily based on actual experiments. When the connection to the real subjects behaviour was considered \cite{kahneman}, an entire host of puzzles and paradoxes appeared even in the simplest artificial (laboratory) conditions. Thus the inclusion of real trader behaviour in the next generation of models and simulations is hampered by the inexistence of comprehensive, systematic, reliable data. Given the present state of the art in psychological experiments, where even the behaviour of single subjects is difficult to assess, we are lead to look for alternative ways to elicit the necessary input for agent-based market modelling.

In this paper we propose a way out of this impasse. Rather than considering the computer as a passive receiver of the behavioural information elicited by psychological experiments, we use the computer itself as an instrument to extract some of the missing information. More precisely, we ask the subjects to write and update  adaptively, between simulation runs (or virtual trading sessions) their own avatars. By gradual corrections, those avatars converge to satisfactory representations of the subjects' behaviour, in situations created by their own collective co-evolution. The fact that the co-evolution takes place through the intermediary of the avatars interaction provides an objective detailed documentation of the process.

More important, the dialogue with the avatars, their actions and their collective consequences assist the subjects in expressing in a more and more precise way their take on the evolving situation and validate the avatar as an expression of the subject internal decision mechanisms. Ultimately, the avatar becomes the objective repository of the subject’ decision making process. Thus we extend, with the help of computers, the behaviorist realm of objectivity to a new area of decision making dynamics. The classical behaviourism limits legitimate research access to external overt behaviour, restraining its scope to the external effects produced by a putative mental dynamics. The method above enables us to study the subjects decision making dynamics without relying on ambiguous records of overt subjects behaviour nor on subjective introspective records of their mental state and motivations.

Far from invalidating the psychological experimental framework, the present method offers psychological experiments a wide new source of information in probing humans mind. The competitive ego-engaging character of the realistic NatLab market platform \cite{NatLab} puts humans in very interesting, authentic and revealing situations in a well controlled and documented environment. Thus standard psychological techniques can exploit it e.g. by interviewing the subjects before and after their updated strategies are applied and succeed (or fail!).

\section{Avatars}

The program sketched in the previous section suggests a Behavioural Finance viewpoint, in a realistic simulation framework. More precisely, the avatars acting in such an environment are able to elicit from the subjects operationally precise and arbitrarily refined descriptions of their decision processes.
In particular, by analysing the successive avatar versions that passed the validation of its owner, one can learn how the owner behaves in this market environment, how (s)he designs his/her strategies, how (s)he decides to depart from them, how (s)he updates them iteratively, etc. Thus the new environment acquires a mixed computational \emph{and} experimental laboratory character. In this respect, the present study owes to previous research that involved simulations / experiments combining human beings and artificial agents, in real-time \cite{cappellini1} or off-line \cite{muchnik2,boer} -- see \cite{duffy} for a review of computational vs experimental laboratories.

The heart of the new simulation-experimentation platform is the co-evolving set of Avatars. They constitute both the interacting actors \emph{and} the medium for recording the chronicles of the emergent collective dynamics \emph{of the subjects}. As a medium for capturing cognitive behaviour, the avatars help extend the behaviorist objectivity criteria to processes that until now would be considered as off-limits. We are achieving it by trying to elicit from humans operational instructions for reaching decisions that they want implemented by their market representatives - the avatars. There is an important twist in this procedure: we are not trying to obtain from the subjects reports of their internal state of mind and its evolution; we are just eliciting instructions for objective actions in specific circumstances. They are however formulated in terms of conditional clauses that capture users intentionality, evaluations, preferences and internal logics.

\subsection{Principle}

\begin{figure}
\centering
\includegraphics[width=8cm]{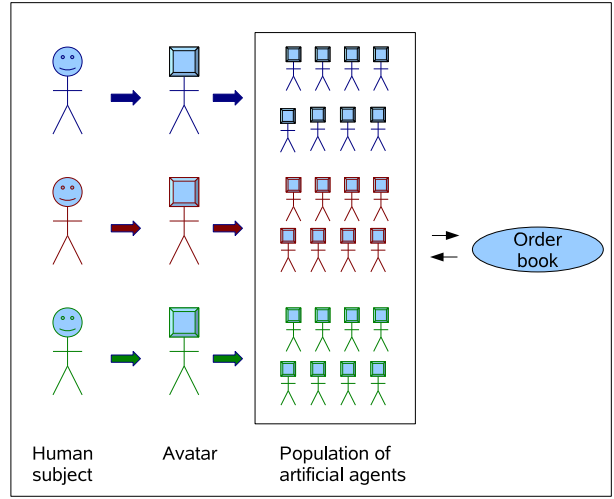}
\caption{The Avatar-Based Method: Human subjects design their avatar in the NatLab environment. From each avatar, a family of artificial agents is generated and included in the market.}
\label{fig:avatars}
\end{figure}

At the beginning of a run, every participant designs his own avatar which is used as a basis to generate an entire family of artificial agents whose individuality is expressed by various (may be stochastically generated) values of their parameters. The resulting set of artificial agents compete against each other in our market environment; see Fig. \ref{fig:avatars}. We use many instances, rather than a single instance of the avatar for each subject, for the following reasons: 
\begin{itemize}
\item having a realistic number of traders that carry a certain strategy, trading policy or behaviour profile
\item having enough statistics on the performance of each avatar and information on the actual distribution of this performance
\end{itemize} 

\begin{figure}
\centering
\includegraphics[width=8cm]{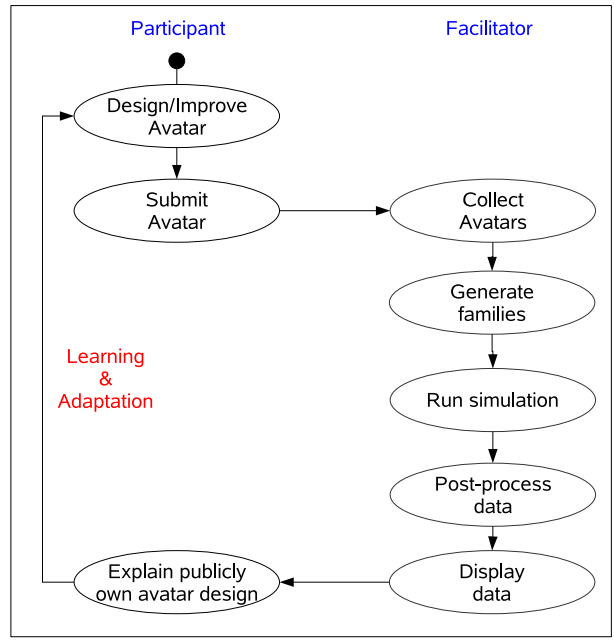}
\caption{Iterative process: participants design and improve their avatar in between every simulation run}
\label{fig:simulationRun}
\end{figure}

Once the population of agents is generated, a first simulation run is performed. A typical run lasts about 10 minutes of CPU , which may represent years of trading on the artificial time scale. At the end of each run, the results are processed and presented to the participants. In our experiments until now, both private and public information were made available. In particular, the price (and volume) trajectory, the (relative) individuals wealth in terms of cash holdings, stock holdings, and their evolution were publicly displayed. The avatar codes were also disclosed and the participants were asked to describe publicly their strategy and the design of their avatar. After being presented with the results (whether full or only public information) of the previous run, the participants are allowed to modify their own avatar and submit an upgraded version for the next run, as described in Fig. \ref{fig:simulationRun}.

The goal of this iterative process, co-evolving subjects thinking with computer simulations, is to converge in two respects; the subject understands better and better:
\begin{itemize}
\item the consequences of his/her own strategy
\item how to get the avatars to execute it faithfully
\end{itemize}

\subsection{Comparison between approaches}

\begin{figure}
\centering
\includegraphics[width=11cm]{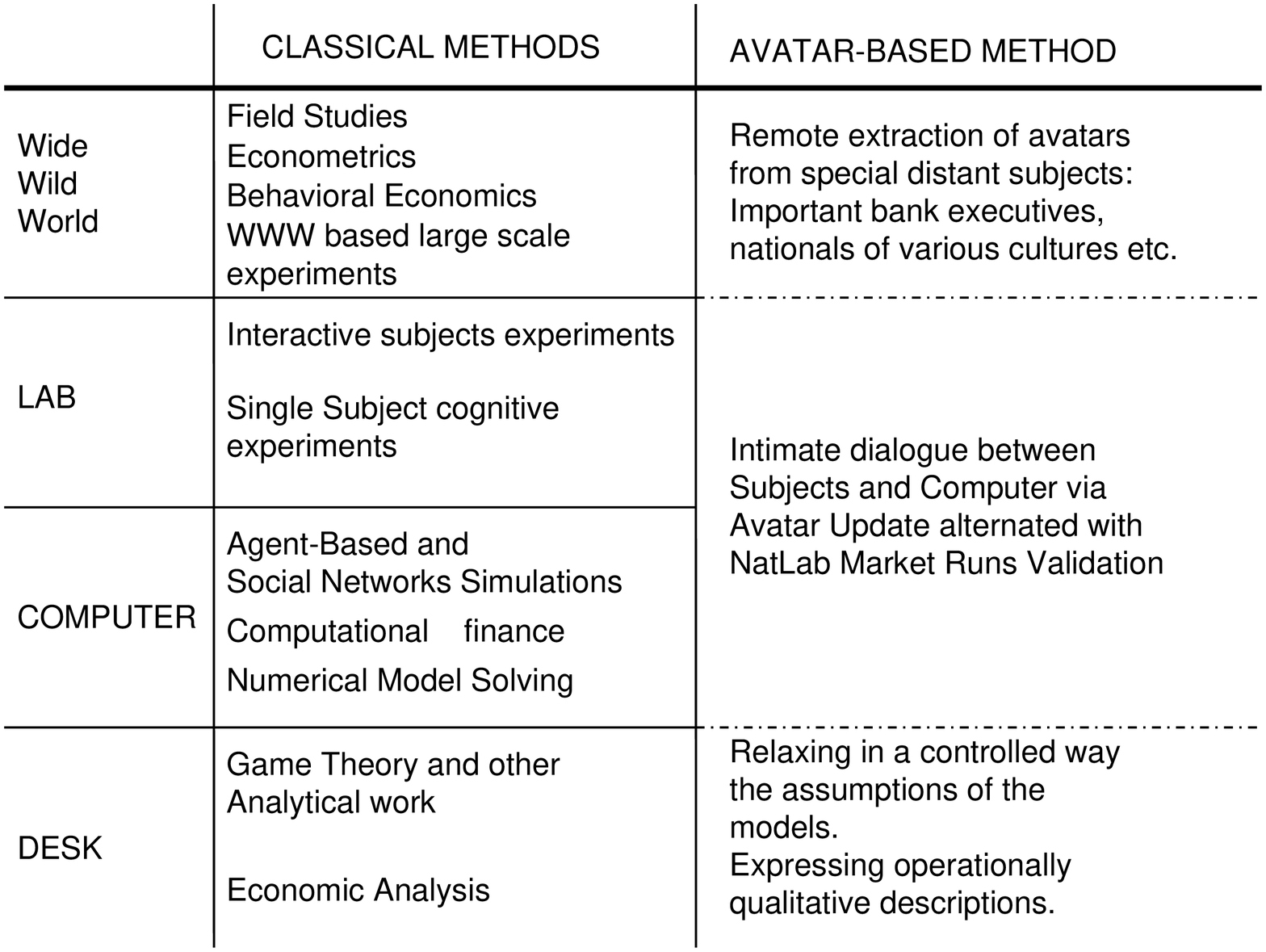}
\caption{Positioning the Avatar-Based Method}
\label{fig:comparisonTable}
\end{figure}

In this section, we discuss the relevance of our method in the context of other works in economics. The economics field spans a wide range of fields and approaches. In the table displayed in Fig. \ref{fig:comparisonTable}, the four rows classify the activities in terms of their context and environment, starting with the DESK at the bottom of the table, extending it to the use of computers, then to the laboratory and ultimately to the real unstructured world.

The two columns of the table refere to the usual methods and the "Avatar-Based Method" (ABM from now on). One sees that in our view, the ABM constitutes a rather uniform way to treat economic behaviour and is capable of bridging application areas that were until now disjoint. This is clearly the case for the LAB and COMPUTER rows, where we even erased the separation line, but it has implications for the other rows too. For instance the Avatars, and especially within the NatLab environment, have been used already to extend to more realistic conditions some theoretical models (row 4 of the table) and results \cite{slanina1}.

At the other extreme (rows 1-2 in the table), the Avatar-Based Method can help correct a perennial problem of economic studies: the biased and sometimes unrepresentative profile of the involved subjects. Indeed, it is very difficult to involve in those studies decision making officials from financial institutions or traders. Substituting them by BA Undergrads is hardly a step towards realistic emulation of the real world. It is much more likely that these important players, rather than coming to a lab, will agree to provide the elements for creating their Avatars. Similar problems can be solved by including in the ABM experiments subjects from far away cultures or environments, without the necessity for distant travels and without separating them from their usual motivations and environment. Moreover, the information provided once by such subjects that are difficult to access can be used now repeatedly by playing their Avatars. Thus ABM has  a good chance to bridge the gap between field studies and lab experiments too (rows 1-2 in the table). In fact as opposed to experiments that do not involve a market mechanism with capital gain and loss, in NatLab, incompetent non-representative subjects will naturally be eliminated since their Avatars loose very quickly their capital.

Another point on which the ABM procedures are offering new hope is the well known problem of subjects motivation. Within the usual experimental frameworks, it is very difficult to motivate subjects, especially competent important people. From our experience, the NatLab realistic framework and the direct identification of the subjects with their Avatars successes and failures, lead to a very intensive and enthusiastic participation of the subjects even for experiments that last for a few days. In fact, beyond the question of "prestige", even seasoned professionals reported to have gained new insights in their own thinking during the sessions. Another promise that ABM is yet to deliver is that by isolating and documenting the Avatar update at discrete times, one will be able to contribute to neighbouring cognitive fields such as learning.

\section{Method Validation}

A piece of software is not an experimental set-up. With all its power, the value of the platform and of the 
"Avatar-Based Experiments" method has to be realized in real life and an elaborate technical and procedural set-up has to be created. The basic condition for the very applicability of our method is the humans capability to faithfully, precisely and consistently express their decision making in terms of computer feedable procedures. Thus we concentrated our first validation efforts in this direction, adapting platform and procedural features to accommodate humans. Many other experimental aspects have to be standardised and calibrated, but in those experiments we concentrated on this crucial \emph{sine qua non} issue. We can conclude at this stage that while there are humans (even economist professionals and successful traders) that could not express their "system" in a computer feedable format ("buy low , sell high"), by-and-large the participants in our experiments were able to confirm at some point that their avatar behaved in agreement with their own behaviour. This happened even with subjects with no particular computer (or market) skills.

\subsection{Experimental setup}
Our experiment features a continuous double-auction implemented on the \emph{NatLab} simulation platform. Every participant received extensive support from a computer scientist to implement his/her avatar in C++ on the platform.

\paragraph{NatLab platform}
The NatLab has the capability to simulate in great detail the  continuum time asynchronous real world \cite{muchnik2}. Bilateral and multilateral communication between agents outside and in parallel with the market is made possible by NatLab. However, given that this experiment focuses mainly on the participants behaviour, we kept the market mechanism (the \emph{rules of the game}) as simple as possible, while retaining the concept of continuous double-auction, essential to understand the price formation dynamics. NatLab was initially engineered as a simulation platform but its use is now in three distinct directions:
\begin{enumerate}
\item the platform provides a realistic framework for the individuals to act within. Providing this "reality" is independent of whether one is interested in its characteristics; it just allows an interactive continuous extraction of information from each of the participants and thereby refining our understanding on their approach, reactions and decision mechanisms;
\item the platform is part of a recent wide effort to understand the emergence of collective complex dynamics out of interacting agents with well defined, relatively simple individual behaviour; and
\item the platform, due to its realistic features and its asynchronous continuous time microstructure, is a reliable way to reproduce and maybe in the future predict real market behaviour.
\end{enumerate}

\paragraph{Market microstructure}
Our market implements a continuous double-auction mechanism, where agents can submit, asynchronously and at any time, limit or market orders to a single public book. Orders are sorted by price and then by time, as on the \emph{NYSE} for instance. Every agent acts as a simple trader, and we do not include brokers or market makers at this stage. In this simple setup, agents balance their portfolio between a risky asset (a stock distributing or not a dividend) and a riskless one (a bond yielding or not an interest rate). Agents can communicate with each other through pairwise messages, and react to external news according to an idiosyncratic sensibility.

\paragraph{Avatars}
We organise our experiment as a competition between participants through the intermediary of their avatars. Avatars generate, by assigning values to their parameters, families of agents that act as independent (but possibly interacting) individuals in the market. The subjects' aim in each run is to generate a family of artificial agents that perform well against other families throughout the simulation run. A typical simulation run is exhibited in Fig. \ref{fig:postProcess}. Families were compared by their average wealth, but an average utility (given some utility function) or a certain bonus for minimising risk could be used in the future.

We give our participants total liberty while implementing their avatar. They can define their own time horizon and design trading strategies as simple or complex as needed, but in the future we may tax agents with heavy data processing by imposing a fine or a specific time lag in the order execution.

\subsection{Preliminary results}
\begin{figure}
\centering
\includegraphics[width=11cm]{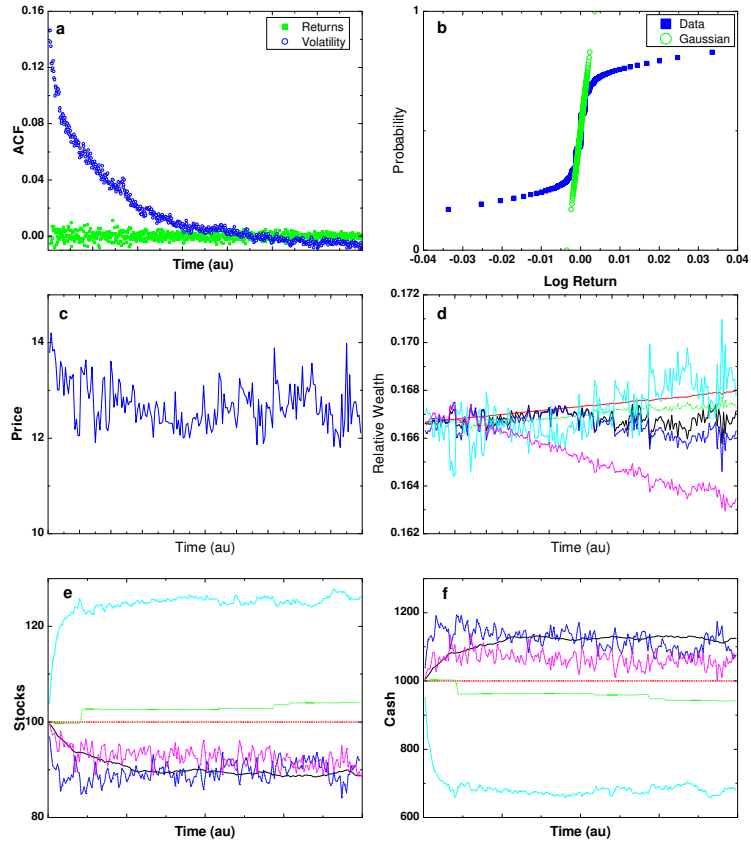}
\caption{Typical run with 7 avatars, 1000 agents each, for above 350 000 transaction ticks. (a) Autocorrelation functions -- absence of raw returns autocorrelation and long-term autocorrelation of volatility, as defined as absolute returns, as observed in empirical data \cite{stanley1};(b) Normality Plot -- fat tailed distribution of returns; (c) Price trajectory; (d) Relative wealth of agents populations -- measure the relative success of competing avatars; (e) Stock holdings -- some strategies are clearly buy-and-hold, others interact with each other; and (f) Cash holdings}
\label{fig:postProcess}
\end{figure}

We have run two sets of experiments so far, with different participants including practitioners (real traders) and academics, either economists, physicists, psychologists or computer scientists. Each experiment included seven participants. The first experiment took place on July 19-31 2004, in Lyon, during the SCSHS Summer School on \emph{Models for Complex Systems in Human and Social Sciences} organised by the Ecole Normale Supérieure de Lyon. The second was organised on January 12-16, in Turin, during the \emph{Winter Seminar on Agent-based Models of Financial Markets} organised by the ISI Foundation. A typical run, with a preliminary analysis of the price time series and relative evolution of populations, is presented on Fig. \ref{fig:postProcess}. We report here on some of the non trivial aspects of the participants behaviour during the experiments, while creating and updating their avatars.

\paragraph{Imprinting oneself}
We noticed, specially at the beginning of the process, that some of our participants encountered some difficulties to express themselves in terms of computer feedable strategies. However, this improved dramatically during the iterative process itself. This is clearly linked to the learning process that one has to face while performing any experiment, especially computerised ones.

\paragraph{Conscious / Unconscious decisions}
The very nature of our method barely allows such things as intuition, improvisation or unconscious decisions to be operationally expressed in the avatar. In fact, after a few runs, avatars capture exclusively the conscious part of our subjects decision making process. Since we we do not know to what extent markets dynamics are driven by unconscious choices, it would be interesting to design a double experiment, comparing subjects and their own avatar in the same market microstructure.

\paragraph{Convergence}
There are two different but related convergence processes that took place during the successive iterations: the first was the convergence of the avatar's behaviour to its creator's intended strategy, while the second involved the evolution of subjects strategy itself to beat other participants. While it appeared relatively easy after a couple of runs to get an avatar successfully reproducing their initial intended behaviour, subjects, driven by competition, kept refining and complexifying their strategy.

\paragraph{Strategies}
An interesting panel of strategies was proposed and grown by the participants, that could loosely be termed random trader, momentum trader, oscillatory trader, diversified Bollinger Bands trader, volume seeker, Neural Network based trader and evolutionary trader. Practitioners clearly outmarked themselves by their ability to think out of the box, the creativity of their strategies, their high analysis power and ability to quickly understand what was going on and spot opportunities to arbitrage other participants' strategies. We also observed the emergence of cooperation between participants to hunt for the leader, trying to bring down the winning strategy by copying and modifying it or even custom-designing new strategies for this specific purpose.

\paragraph{Fundamental Value}
In the two experiments we ran, our computer simulation figured a closed artificial market, with no creation of stocks, no distribution of dividends and no interest rate associated with the riskless asset, cash. In those conditions, we observed that after a transition period, characterised by high volumes, during which assets were heavily reallocated between agents, the price kept fluctuating around a steady state equilibrium price. This price, emerging from the interactions between heterogeneous, relative risk aversion agents, was generally different from the fundamental value we could have expected from rational agents with homogeneous preferences.

\section{Conclusion}

The rapidly growing field of Agent-based Computational Finance comes naturally as a complementary approach to the other Finance subfields: Behavioural Finance, Laboratory experiments, Econometrics, Game Theory, etc. The field is definitely out of his infancy and a rather wide range of choices is available to academics and practitioners that wish to define and test concrete real and realistic systems or new models of individual and market behaviour. The next step is to set common standards for the platforms that propose to represent and simulate artificial financial markets \cite{markowitz1,cappellini,muchnik1,boer}. One possible goal is to transform them in virtual or even real laboratories capable to implement and test in realistic conditions arbitrarily sophisticated experiments. One way to solve the problems of realistic trader behaviour is the Avatar-Based Method introduced in the present paper. Even though there are many obstacles not even yet uncovered in realizing its ambitions, the method is already providing new insight and definitely even if its main ambitions are going to remain unfulfilled, it is guaranteed to provide fresh unexpected and valuable material to the existing methods.

Among the fundamental issues which the ABM can address is the mystery of price formation by providing in great detail, reliability and reproducibility, the traders decision making mechanisms. Occasionally the Avatars are going to be caught unprepared and inadequate to deal with some instances that were not previewed by their owners. By the virtue of this very instance, they will become effective labels for the emergence of novelty in the market. Thus in such instances, even in its failure, the ABM will provide precious behavioural and conceptual information.

ABM can serve as a design tool for practitioners in the development of new trading strategies and the design of trading automata. Moreover, we hope that this approach will provide new ways to address some of the fundamental problems underlying the economics field:
\begin{itemize}
\item how people depart from rationality
\item how out-of-equilibrium markets achieve or not efficiency
\item how extreme events due to a shifting composition of markets participants could be anticipated
\end{itemize}

The experiments we ran, beyond eliciting information, provided a very special and novel framework of interaction between practitioners and academics. Thus NatLab and ABM might have an impact on the community by providing a common language and vocabulary to bring together academics and much needed practitioners. As a consequence, it appears necessary to gather interdisciplinary projects that would house within the same team the psychologists that run experiments on people's behaviour, computer scientists that canonise this behaviour into artificial agents, practitioners that relate those experiments to real markets and economists that assess the consequences in terms of policy making.

\section*{Acknowledgements}
We would like to thank the participants of our experiments for their time and commitment, together with the participants of the Seminar on \emph{(Un)Realistic Simulations of Financial Markets} at ISI Foundation, Turin, Italy, on April 1-5 2005, for their enlightening comments, from which this paper largely benefitted. Finally, we are largely indebted to Alessandro Cappellini and Pietro Terna for their views and experience on online laboratory experiments of stock markets, as well as Martin Hosnisch, Diana Mangalagiu and Tom Erez. The research of SS was supported in part by a grant from the Israeli Academy of Science, and the research of LM was supported by a grant from the Centre for Complexity Science. All errors are our own responsibilities.

\bibliographystyle{alpha}
\bibliography{bibliography}
\end{document}